# Neural Network Segmentation of Interstitial Fibrosis, Tubular Atrophy, and Glomerulosclerosis in Renal Biopsies

Brandon Ginley[1,†], Kuang-Yu Jen[2,†], Avi Rosenberg[3], Felicia Yen[2], Sanjay Jain[4], Agnes Fogo[5], Pinaki Sarder[1,6,7,*]

[1]Departments of Pathology & Anatomical Sciences, [6]Biostatistics, and [7]Biomedical Engineering
University at Buffalo – The State University of New York, Buffalo, New York

[2]Department of Pathology and Laboratory Medicine, University of California, Davis Medical Center, Sacramento, California

[3]Department of Pathology, Johns Hopkins University School of Medicine, Baltimore, Maryland

[4]Division of Nephrology, Department of Medicine, Washington University School of Medicine, St. Louis, Missouri

[5]Departments of Pathology, Microbiology, & Immunology and Medicine, Vanderbilt University, Nashville, Tennessee

[†]Indicates equal contribution

[*]Address all correspondence to: Pinaki Sarder
Tel: 716-829-2265; E-mail: pinakisa@buffalo.edu

**Significance statement** Transplantation of the kidney is the most effective strategy for treating end-stage kidney disease, and results in significant improvements to quality of life and survival as compared to dialysis. A fundamental component to selecting and monitoring allograft tissue is the kidney biopsy, and biopsies may be taken both pre- and post-operatively to determine the functional health of the tissue based on histological structure. Histopathological analysis of biopsies provides valuable information regarding presence of acute rejection or other potentially reversible tissue injuries of the transplant patient as well as the amount of irreparable allograft injury. However, this analysis is based on the observing pathologist's visual identification and interpretation of structure, which are subject to inter and intra-rater reliability. Computational evaluation of histological biopsy data can improve allograft evaluation by providing rapid, precise, unified interpretation of structure, thereby increasing the amount and reproducibility of clinical information.

## ABSTRACT

**Background:** Glomerulosclerosis, interstitial fibrosis, and tubular atrophy (IFTA) are histologic indicators of irrecoverable kidney injury. Cortical fibrosis alone is the most robust morphologic predictor of chronic kidney disease across the entire spectrum of kidney disease. The extent of biopsy tissue demonstrating these histomorphologic parameters is an estimate of remaining functional renal mass and are indicators of renal prognosis, independent of disease etiology. In standard clinical practice, the renal pathologist visually assesses, under the microscope, the percentage of sclerotic glomeruli and the percentage of renal cortical involvement by IFTA. Although percentage of glomerulosclerosis is fairly straightforward and shows low inter-observer variability, visual quantitation of IFTA presents a significantly greater challenge due to its more complex histologic definition and varied spectrum of morphologic manifestations. Modern artificial intelligence and hardware have enabled identification of arbitrary image structures with enhanced precision and speed. Accurate quantitation of morphologic features can also be achieved through morphometric analysis. Such advantages can be leveraged to possibly improve diagnostic and prognostic capabilities based on biopsy information. Here, we applied convolutional neural networks (CNNs) to semantically segment glomerulosclerosis and IFTA in renal biopsies.

**Methods:** Periodic acid-Schiff-stained whole slide images of 65 renal biopsies (transplant biopsies with varying levels of IFTA, native kidney diabetic nephropathy, and control) were annotated by a renal pathologist for non-sclerotic glomeruli, sclerotic glomeruli, and IFTA, of which 38 cases were used to train various configurations of CNNs. CNN configurations included using a smaller versus a larger network and analyzing the images at low resolution, high resolution, or a combination of low/high resolution. The training set included transplant cases and native cases (with diabetic nephropathy) processed and stained at two different institutions. The testing set consisted of 14 holdout cases from the same two institutions that the CNNs were trained on (intra-institutional holdout cases) as well as a set of 13 cases from a third institution that the CNNs were not trained on (inter-institutional holdout cases). The CNNs' performance for identifying non-sclerotic glomeruli, sclerotic glomeruli, and IFTA was tested on both intra-institutional and inter-institutional holdout cases. CNN performance was based on comparison to ground truth annotations by the pathologist.

**Results:** The CNN showing the best performance was based on a smaller network analyzing the images at low resolution. This CNN demonstrated a balanced accuracy of $0.82/0.94/0.86$ and MCC of 0.6 / 0.87 / 0.68 on intra-institutional holdout cases and balanced accuracy of 0.70 / 0.93 / 0.84 and MCC of 0.49 / 0.79 / 0.64 for inter-institutional holdout cases for the identification of IFTA, non-sclerotic glomeruli, and sclerotic glomeruli, respectively. Moreover, in many cases, the network learned to predict IFTA boundaries more precisely than the rough annotations used for training. Estimations of IFTA and glomerulosclerosis were substantially correlated with the ground truth annotations, showing correlation coefficients of 0.73 (95% confidence interval (CI) [0.31, 0.91]) and 0.97 (95% CI [0.9, 0.99]), respectively. When comparing IFTA grades (absent, mild, moderate, severe) based on the pathologist's visual assessment and CNN predictions against ground truth annotations, there was not a substantial difference in score agreement (kappa 0.69, 95% CI [0.39, 0.99] for pathologist; kappa 0.66, 95% CI [0.37, 0.96] for CNN). Interestingly, the CNN was able to learn to identify segmental sclerosis without having been trained on such histopathologic findings.

**Conclusions:** CNN segmentation of morphologically complex image structures such as IFTA is feasible and can show high performance. Further improvement in performance to overcome certain limitations can likely be accomplished through further expansion of the training set on biopsies with varied histologic manifestations of IFTA.

1. **INTRODUCTION**

Histopathologic changes observed in kidney biopsies are not only vital for the diagnosis of kidney disease, but also serve as valuable prognostic and predictive markers. The primary histologic indicators of irreparable renal injury include moderate to advanced glomerulosclerosis and interstitial fibrosis and tubular atrophy (IFTA). These key morphologic features of chronicity are thought to estimate the overall degree of irrecoverable renal mass. IFTA is the best morphologic prognostic marker for disease progression and renal outcome, irrespective of the etiology of the renal disease[1-12]. Furthermore, patients with advanced chronicity are unlikely to recover any significant amount of renal function from treatment. The degree of chronicity is more important in determining treatment strategy, potentially sparing patients with advanced chronicity from toxicities of various medications.

IFTA is subject to more variability between pathologists than glomerulosclerosis[13-15]. IFTA is assessed based on visual estimation of overall percent renal cortical area involvement, rather than counting and quantifying discreet, individual structures, as is the case for glomeruli. Given that IFTA typically occurs in a patchy fashion, visual estimation can be difficult as the pathologist tries to mentally "add up" all the areas of IFTA. This process leaves room for error in the pathologist's assessment and further complicates inter-observer reliability. The concurrent examination of periodic acid-Schiff (PAS) and trichrome-stained sections in addition to the standard hematoxylin and eosin (H&E) stain can increase the pathologist's sensitivity and specificity for IFTA detection, thereby improving accuracy and reducing inter-pathologist variability. Furthermore, the most widely used grading scheme for IFTA helps to reduce inter-pathologist variability since it is based on "round" cut-off values for mild ($\leq$25%), moderate ($\leq$50%), and severe (>50%). However, a major consequence is the large range of percent IFTA that can be categorized into the same grade, thus possibility reducing the utility of IFTA as a prognostic and predictive indicator. In other studies, rounding to nearest 10% has been used, and showed good correlation with morphometry based assessment. Although pathologist "eye-ball" assessment of IFTA has shown reasonable reproducibility compared to morphometry, we sought to further enhance reproducibility and accuracy by applying neural network learning.

Contemporary digital image analysis routines have shown promise as a way to accurately recognize and measure specific features on biopsies that may be of clinical value. Major advantages of such technology include the ability to at least partially automate biopsy analysis and to provide more precision and accuracy in morphometric qualification of specific histologic findings. This in turn enables the extent of features such as IFTA to be reported as continuous variables rather than traditional categorical variables, which may show improved clinical utility. In this way, the increased depth of the data can be used to build better models and guidelines in the treatment of kidney diseases.

In this study, we sought to automate the identification and quantification of IFTA and glomerulosclerosis percentage on whole slide images (WSIs) of PAS-stained kidney biopsies using a fully convolutional neural network (CNN). We selected the DeepLab V2[16, 17] network as our CNN architecture, because of its prior successes in semantic segmentation tasks[16] and its easy-to-use open source framework[17]. The network models were trained using a modified version of our previously published H-AI-L (Human Artificial Intelligence Loop)[18] pipeline for WSI annotation and convolutional network training. The modifications facilitated efficient management of non-sparse image data at multiple resolutions (low, high, and combined low-high) to compare efficacy in segmentation of non-sparse regions of interest (i.e. IFTA). We found that a CNN approach was highly robust for the detection of IFTA and glomerulosclerosis. Interestingly, the network learned to be more precise in IFTA detection than its training data, developed its own conceptualization of segmental glomerulosclerosis, and was able to detect IFTA and glomerulosclerosis in WSIs from an institution on which it was not trained.

## 2. METHODS

### 2.1. Image data

This study was approved by the Institutional Review Board at the University of California, Davis (UCD) and University of Buffalo. Image data consisted of WSIs from 35 kidney transplant biopsies, 28 biopsies with diabetic nephropathy (DN), and 2 normal kidney controls. Transplant biopsies were selected from the specimen archives at UCD to represent the full spectrum of IFTA severity that is typically encountered in the clinical setting. DN biopsies were selected either from the Kidney Translational Research Center at the Washington University School of Medicine St Louis or from the Vanderbilt University Medical Center. In all cases, tissue sections were prepared at 2 µm thickness and stained with PAS. Slides were scanned using a brightfield microscopy whole slide scanner (Aperio®, Leica) at 40x magnification resulting in an apparent resolution of 0.25 µm/pixel.

### 2.2. Ground truth annotations

Annotation was directly performed on WSIs using our H-AI-L pipeline[18]. This pipeline was designed to ease the annotation burden required for training CNNs on WSIs. WSI viewers like Aperio® ImageScope provide native annotation tools, such as the pen tool, to identify regions of interest at arbitrary resolutions on the entire WSI. H-AI-L acts as an intermediary between ImageScope and the DeepLab V2[16, 17] segmentation network, optimizing the interface between WSIs and CNNs. ImageScope stores whole slide annotations in XML format, where regions are represented by the collection of boundary points which contain them. The H-AI-L pipeline converts these boundary points to image masks for CNN training, and can additionally convert network output back into whole slide annotations. All annotations were provided with the pen tool in ImageScope by a renal pathologist and were considered ground truth in this study. All annotations were performed on PAS-stained WSIs. Adjacent trichrome WSIs were available for the pathologist to review to help with annotation for the transplant biopsies. Three classes were annotated: IFTA, non-sclerotic glomeruli, and sclerotic glomeruli.

## 2.3. Case selection

Because IFTA is a highly complex image structure with a large spectrum of morphologies, we restricted the type of cases used in this pilot study to those with well-established areas of IFTA. Examples of cases with difficult to define IFTA boundaries include those with early or evolving IFTA with variable intermixed edema, those with significant differences between the extent of interstitial fibrosis versus tubular atrophy, and those with heavy inflammatory cell infiltrate. The reason for this decision was that these cases typically do not have consensus among renal pathologists and are in a metaphorical "gray zone". These image structures would likely only obfuscate the network's learned model developed on cases with more evident boundaries. Cases with some medullary sampling were preferentially selected so that it could be validated that the network is able to differentiate between cortex and medulla. This pilot model can be used in future studies as a pre-trained model to investigate more challenging cases.

## 2.4. CNN training and specifications

Segmentation of image regions was performed using the DeepLab V2[17] CNN, built in Tensorflow[19]. Two different network variations were tested: one with a DeepLab encoder and one with a ResNet-50 encoder. The comparison was to analyze whether performance gains could be achieved by using a smaller network (ResNet-50 encoded) and training with larger batch sizes. The DeepLab encoded model was initialized on a checkpoint pretrained on MSCOCO dataset[20]. The ResNet-50 encoded model was initialized on a checkpoint pretrained on the ILSVRC dataset[21].

**Training and testing cohorts**

Input training images were chopped using a sliding window technique from $n=38$ WSIs (25 transplant, 11 DN, and 2 control nephrectomies), and testing images were chopped identically from $n=14$ intra-institutional holdout biopsy WSIs (10 transplant, 4 DN) or from $n=13$ inter-institutional holdout cases (from DN patients). Intra-institutional cases were holdout data that were produced in the same institution as the data that the network was trained on. Inter-institutional cases were holdout data that was produced in a different institution than the data the network was trained on. Fewer DN cases were used for training than transplant cases because fewer were available, and in DN, IFTA was more prominent and less variable (e.g. prominent tubular basement membrane thickening leading to high contrast).

**Network training specifications**

Analysis was compared at two magnifications, one at low resolution (the corresponding 40x magnification image downsampled 4x in each dimension), and one at high resolution (40x magnification). In both the low and high resolution analyses, the field-of-view of the images was set to be 450 pixels in each dimension. DeepLab encoded networks were trained with a batch size of 2, and ResNet-50 encoded networks were trained with batch size 4. Input images were augmented at training time via random flipping vertically and horizontally, random piecewise affine geometric distortion, absolute shifting of the H component of the HSV space, and gamma shifting of the $L^*$ component of the $L^*a^*b^*$ space. The latter two shift amounts were selected from normal random distributions with $\mu=0, \sigma=0.025$ and $\mu=1, \sigma^2=0.05$, respectively, for each augmented image. Number of augmentations for each class was determined by inverting the pixel-wise fraction of each class in the entire training set and scaling it by a maximum number of augmentations, $\alpha$, in this case, $\alpha=15$ for low resolution patches, and $\alpha=3$ for high resolution patches. The high resolution network was trained for one epoch because the sheer size of the dataset made it infeasible to compare multiple network configurations in a timely manner if multiple epochs were specified (i.e. >1 million images requiring approximately 100 hours for each epoch). Furthermore, we found significant training past one epoch only caused the network to begin overfitting the data. To train the low resolution network for a similar amount of time, this network was trained for 16 epochs because there is 16-times less data in the low resolution dataset due to down-sampling. Both networks were trained with initial learning rate $2.5e^{-4}$, power 0.9, momentum 0.9, and weight decay 0.0005.

## 2.5. Comparison of traditional and computational estimation models

Segmentation performances were calculated by pixel-level comparison of entire WSI predictions against manually annotated ground truth WSI annotations. Reported performance metrics include sensitivity, specificity, balanced accuracy[22] (because the classes were extremely imbalanced), and Matthew's correlation coefficient (MCC)[23]. The CNN's estimation of percent renal cortical involvement by IFTA was compared to the pathologist's visual estimation. The pathologist who initially annotated the WSIs was subjected to a washout period of 15 months and was then asked to reevaluate the 14 holdout cases for percentage of IFTA through visual estimation only, without annotating the IFTA borders. Visual estimation was performed to the closest 5%, and the adjacent trichrome WSI was available to the pathologist for the transplant cases. IFTA estimations on WSIs were calculated by dividing the area of detected IFTA by the area of the cortex. The cortical region of each biopsy was marked manually, but only for the purpose of calculating the total cortical area, not for restricting the computer's IFTA predictions. $IFTA^{ti}$ was calculated by calculating the area of detected IFTA minus contained glomerular area, divided by the area of the cortex minus the total glomerular area (including both sclerotic and non-sclerotic glomeruli). Glomerulosclerosis index was calculated by dividing the number of sclerotic glomeruli by the total number of glomeruli.

A single case had deep medulla that displayed extensively thickened tubular basement membranes and hyaline casts. This morphology of the medulla was not encountered in the training set and therefore caused the CNN to erroneously identify these medullary regions as cortex involved by IFTA (of note, the CNN was trained *not* to predict IFTA in medullary regions). As a result, the amount of IFTA was vastly overestimated by the CNN on this case, which was a statistical outlier (>3 standard deviations greater than the mean). Thus, this case was excluded from correlation calculations.

We expect that simply including more of this morphologic presentation in the training data would eliminate this issue, because the CNN did not predict significant amounts of IFTA in the medulla regions on the other 13 holdout cases. Recent works[24] have rigorously demonstrated that the greatest barrier to a generalized deep learner is the number of different cases. Further, even more than 40,000 whole slide images may not allow the learner to be fully generalized. These numbers were not possible to acquire in our study, however, we are still able to achieve substantial correlation with the manual based method.

## 3. RESULTS

### 3.1. H-AI-L pipeline overview

The goal of this study was to explore the feasibility of using CNNs to automatically identify and quantify regions of IFTA within the renal cortex, and to enumerate sclerotic and non-sclerotic glomeruli. As with any supervised machine learning technique, a gold standard or ground truth data set must be used to train the algorithm. To facilitate WSI annotation and subsequent convolutional processing on WSIs, we used our previously published, publicly available H-AI-L pipeline, which consists of a library of functions that allows investigators to train CNNs directly from annotations made within the popular WSI viewer Aperio ImageScope®.

Figure 1 outlines this study's approach using H-AI-L. First, a renal pathologist directly annotated a training set of biopsies on the WSIs by marking the desired regions of interest with a circle in the image viewer (Figure 1A). Multiple annotation layers were used to annotate separate classes of objects including IFTA, non-sclerotic glomeruli, and sclerotic glomeruli. Next, H-AI-L chopped the WSI into overlapping image patches and generated labeled mask images using the boundaries provided by the annotations (Figure 1B). These patches and masks were subsequently used to train a CNN. The trained CNN was then tested for its ability to identify the same type of morphologic features on a new set of image patches (Figures 1C and 1D). The test patches and masks were stitched back together into a WSI mask (Figure 1E), which was then converted to XML format based on region boundary points. The output XML annotation file based on CNN predictions can be read and displayed by the whole slide viewer (Figure 1F).

### 3.2. Segmentation of IFTA

**Design strategy**

Because IFTA is a highly complex histopathologic feature that can manifest as a spectrum of morphologies, creating a viable computational model to define its boundaries required several important considerations in study design. First, case selection for creating a training set based on ground truth annotations was restricted to cases that contained "well-developed" IFTA. Such cases showed mostly classic morphologic features of IFTA characterized by dense collagen accumulation between tubules, i.e. interstitial fibrosis, and decreased tubular circumference with thickened and wrinkled tubular basement membranes, i.e. tubular atrophy. Areas that showed "thyroidization" and "endocrinization" of the tubules were included, although these types of tubular atrophy were infrequently represented. Cases which were deemed "difficult" for IFTA interpretation were excluded since they typically demonstrate poor concordance between pathologists. Examples of such cases were those that showed early or evolving IFTA with variable intermixed edema, those with significant differences between the extent of interstitial fibrosis versus tubular atrophy, and those with heavy inflammatory cell infiltrate.

Second, even when IFTA is "well-developed", there is no quantifiable definition or biological marker that can be used to clearly delineate its precise borders. Thus, the most viable and practical approach was to base ground truth on annotations of a domain expert (i.e. renal pathologist). Given that in clinical practice, IFTA grading is performed as a visual estimation of the percentage of renal cortical involvement without delineating precise borders, annotations were performed by examining the WSIs at low power and drawing "rough" borders around IFTA regions. Thus, inclusion of some non-atrophic tubules and small patches that may not represent "well-developed" IFTA did occur. However, this procedure was more reflective of clinical practice and more time efficient.

The final major design consideration was determining the ideal magnification level (i.e. resolution) for analysis, and relatedly, the amount of contextual information provided to the network in each image patch. IFTA is typically evaluated at low magnification by the pathologist, or by estimating an average from multiple medium magnification fields. Similarly in our work we tested both low and high magnification levels for training of the CNN. Incorporating large amounts of contextual information was achievable only at low resolution analysis and was important for the network for two major reasons: 1) to differentiate cortex from medulla and 2) to recognize contiguous regions of IFTA. When the magnification of training image patches was too high (i.e. high resolution and too "zoomed in"), the network did not have enough contextual information to determine whether the region of interest was in the cortex or medulla. As a consequence, the amount of predicted IFTA by the CNN was erroneously elevated, presumably due to the morphologic features of the medulla that could resemble IFTA (supplementary Figure 1). Also, when the analysis was performed at high resolution, the network would have difficulty connecting closely clustered IFTA regions that were clearly one contiguous region of IFTA when viewed from a low magnification perspective (Figure 2).

**Performance**

Various CNN configurations were developed and tested, which involved 1) using either the DeepLab encoder or the ResNet-50 encoder and 2) performing the analysis at low resolution, high resolution, or combined low/high resolution. The performance for these CNNs in terms of sensitivity, specificity, balanced accuracy[22], and Matthews correlation coefficient (MCC)[23] is shown in Table 1. Sensitivity and specificity are two of the most commonly reported binary performance metrics but may be misleading if classes are highly imbalanced. Balanced accuracy is an arithmetic average of sensitivity and specificity; and gives an estimation of what the accuracy would be if the classes were perfectly balanced. MCC is a measure of the overall correlation between the predicted and known pixel classification labels, and ranges between -1 (perfect pixel-wise disagreement with the annotator), 0 (no better than random chance), and +1 (perfect pixel-wise agreement). MCC is a superior binary performance metric because it accounts for all true positives, false positives, true negatives, and false negatives. It is also reliable when classes are imbalanced. Thus, in our opinion, MCC is the optimal performance metric for analysis of IFTA results in this study, as conceptually, it reflects the degree of visual overlap between the CNN predictions and the ground truth annotations.

An example of a case with low MCC of 0.51 is illustrated in Figures 3A and 3B. The pathologist's annotation (Figure 3A) and the CNN prediction (Figure 3B) are in the same approximate locations, but the latter boundaries are patchy and sparser compared to the annotation. Additionally, the network included a large region that the pathologist deemed to be inflamed perivascular stroma adjacent to a large caliber vessel rather than IFTA. In contrast, Figures 3C and 3D demonstrate an example of a case with high MCC of 0.84. The annotation (Figure 3C) and predicted boundaries (Figure 3D) very precisely overlap.

In terms of IFTA, all ResNet-50-encoded networks performed better than their counterpart DeepLab-encoded networks as measured by sensitivity, balanced accuracy, and MCC, while specificity was comparable (Table 1). The smaller networks encoded by ResNet-50 allowed more training images to be batched at one time (4 versus 2), which resulted in superior CNN performance. Generally, low resolution analyses outperformed high and combined resolution approaches in terms of MCC. Performing the analysis at low resolution (16x total reduction in magnification) allowed a much greater amount of contextual information (proportionally 16x more) to be included in each image patch, which resulted in more properly contiguous predictions (Figure 2). An extra benefit of low resolution analysis was that training and testing took significantly less time (proportionally approximately 16x less time). Overall, the ResNet-50-encoded network performing IFTA analysis at low resolution appeared to stand out as the best performing network, achieving the highest MCC (0.6) and balanced accuracy (0.82).

Because there were more transplant cases in the training set, the CNN would be expected to perform better on transplant biopsies. However, when intra-institutional holdout cases were grouped based on either DN or transplant biopsies, the optimal low resolution ResNet-50-encoded network identified IFTA equally well in both disease types

(mean MCC of 0.6 for both). The reason for this unexpected finding may be due to less variable morphology of IFTA and more prominent tubular basement membrane thickening seen in DN biopsies.

In Table 1, many of the sensitivities are lower than what might be expected for an ideal segmentation algorithm. Two main factors caused the lower than expected sensitivities. First, the training set size used in this study was somewhat limited. It is well-recognized that neural network approaches generally benefit from large data set sizes. Although the optimal number of cases for IFTA detection cannot be precisely determined, it is likely to be in the tens of thousands[24], which was unfeasible for our study.

The second main factor in causing lower than expected sensitivities is due to the imprecise annotations of IFTA. Since IFTA often occurs without precise borders and can be intermixed with unscarred renal cortex, it was impractical for the pathologist to annotate IFTA at the level of individual tubules and their surrounding interstitium. Instead, the pathologist's annotations were drawn around contiguous regions of what was deemed to be IFTA, which included single or small groups of scattered non-atrophic tubules and small regions without clear interstitial fibrosis. This protocol was used since it more or less replicates the typical renal pathologist's approach to visual estimation of IFTA in standard clinical practice. Thus, the ground truth (which consisted of the pathologist's annotations) included small regions of unscarred renal cortex within areas annotated as IFTA. Although these imprecise annotations were used for training, the CNNs were able to learn the correct morphologic appearance of IFTA and draw more precise boundaries than the initial annotations. The boundaries were so precise that they outperformed what can be provided reasonably by manual annotation. For example, Figure 4A shows the annotation drawn by the pathologist for IFTA as the green border, which includes a large region deemed to be affected by IFTA. This same field as predicted by the CNN (Figure 4B) shows smaller regions predicted as IFTA since the network appears to avoid single and small groups of tubules that show no obvious atrophic changes. The result was an increased pixel-based false-negative rate, which translated to a lower sensitivity.

### 3.3. Glomerular segmentation

Identification of total glomeruli and enumeration of glomerulosclerosis are fairly straightforward, and typically these features show a high level of concordance between pathologists. CNNs have already been shown to be successful for glomerular segmentation in a number of different studies, by us and others[18, 25-27]. Nevertheless, estimation of glomerulosclerosis percentage is a standard part of renal biopsy assessment as it reflects the extent of irrecoverable glomerular injury and along with IFTA are key indicators of overall chronicity. Thus, we included glomerular segmentation with identification of non-sclerotic and sclerotic glomeruli within our CNN model.

Examples of network predictions of glomerulosclerosis by the low resolution ResNet-50-encoded network are shown in Figure 5. Yellow boundaries indicate detected pixels of non-sclerotic glomeruli and red boundaries indicate detected pixels of sclerotic glomeruli. As opposed to the IFTA analysis where the low resolution ResNet-50-encoded network stood out as the best performing network, many of the various CNN configurations showed similar performance (Table 1). As expected, all network configurations demonstrated far superior performance for the identification of non-sclerotic and sclerotic glomeruli over IFTA, given the complex and challenging morphologic definition of IFTA. All network configurations demonstrated higher performance in detecting non-sclerotic glomeruli than sclerotic glomeruli in holdout cases. This result is to be expected because in most of the biopsies there were less sclerotic glomeruli than non-sclerotic glomeruli. Thus, the network had less examples of sclerotic glomeruli to train on, and therefore, it did not identify this class as robustly as non-sclerotic glomeruli.

An intrinsic property of a fully convolutional network is that every pixel is given an individual label, which implies that the network should be capable of predicting segmental glomerulosclerosis. This exact phenomenon was observed in our CNN even though the training data contained no labels of segmental glomerulosclerosis (Figures 5E and 5F).

### 3.4. Intra-institutional versus inter-institutional network predictions

One major limitation of automated computational image analysis algorithms is their inability to adapt to variations in stain characteristics and section quality. The lack of universal staining protocols, differences in reagents from suppliers, variability in tissue sectioning by histology technicians, and a myriad of other possible causes can contribute to variability in final tissue appearance on slides. Furthermore, given the need to digitize slides for computational analysis, variation can be introduced from slide scanning and subsequent digital processing. Automated slide stainers can generate more consistency between batches of slides; however, these methods are generally not consistent across institutions, resulting in widely varying image characteristics for the same tissue type. These visual differences are easy for a human eye to assimilate but are frequently a limiting factor in digital quantification algorithms.

To test the robustness of the CNNs for glomerulosclerosis and IFTA detection, WSIs of PAS-stained sections from two different institutions were included in the training and testing sets and designated as "intra-institutional" cases (Figure 6A and Figure 6B). As a further validation step, renal biopsies sectioned, stained, and imaged at a third institution, which *were not included* in the training set, were tested for glomerulosclerosis and IFTA detection by the CNNs and designated as "inter-institutional" cases (Figure 6C). Even at low magnification as depicted in Figure 6, the staining characteristics are drastically different in color and contrast between the three institution's PAS stains.

As expected, in general the CNNs showed inferior performance for inter-institutional holdout cases when compared to intra-institutional holdout cases. This observation is especially evident for IFTA detection when measured by balanced accuracy and MCC. For example, the low resolution ResNet-50-encoded network balanced accuracy and MCC decreased from 0.82 and 0.6 for intra-institutional holdout cases to 0.7 and 0.49 for inter-institutional holdout cases, respectively. Nevertheless, the metrics for inter-institutional cases still showed fairly high performance. Detection of non-sclerotic glomeruli was fairly comparable between intra-institutional and inter-institutional holdout cases in terms of balanced accuracies, although MCCs were slightly inferior for inter-institutional holdout cases. In contrast, detection of glomerulosclerosis was comparable for both holdout sets for both balanced accuracy and MCC.

These results testify to the extreme robustness of convolutional approaches. The classification configuration which did best on the intra-institutional holdout set also performed the best on the inter-institutional holdout set (as measured by MCC), supporting that the network is somewhat adequately fit and is unlikely to be overfit.

### 3.5. Comparison of visual estimation to network predictions

The renal pathologist who initially annotated the holdout cases for IFTA went back to these cases after a washout period to perform visual assessment of IFTA as performed in standard clinical practice. In particular, the pathologist recorded a visual estimation of percent renal cortical involvement by IFTA to the nearest 5%. Since the low resolution ResNet-50-encoded network showed best overall performance, the pathologist's assessment was compared to the predictions performed by this network. Comparison was also made to morphometrically quantitated measure of these features based on the pathologist's original annotations.

The computational IFTA estimation results for each of the holdout cases is shown in Table 2, compared against the estimations made by pathologist. The visual estimates of IFTA highly correlated with estimates made on annotation boundaries with correlation 0.96 and 95% confidence interval (CI) [0.88, 0.98]. The CNN-predicted IFTA boundaries showed lower (but still moderate to strong) correlation with the annotated boundaries with correlation 0.73 and CI [0.31, 0.91]. In contrast, when the IFTA percentages were converted to IFTA Banff scores, the Cohen's kappa for the visual estimates compared to the annotation boundaries and CNN predicted boundaries were comparable at 0.69 (CI [0.39, 0.98]) versus 0.66 (CI [0.37, 0.96]), respectively.

With regards to the estimation of the percent glomerulosclerosis, the CNN was very accurate with respect to the pathologist, much more than IFTA. This is not surprising as the detection performance for glomeruli and sclerotic glomeruli was much higher than that for IFTA. The CNN-predicted glomerulosclerosis percentage correlated with the pathologist predicted glomerulosclerosis percentage (number sclerotic glomeruli divided by total glomeruli) with coefficient 0.97, 95% CI [0.9, 0.99] (Table 3).

Given that precise areas can be quantitated using the pathologist's annotated boundaries and the CNN's predicted boundaries, we also calculated the tubulointerstitium-restricted IFTA percentage ($IFTA^{ti}$). $IFTA^{ti}$ was defined as percentage of renal cortical involvement by IFTA, excluding all areas occupied by glomeruli. Theoretically and conceptually, $IFTA^{ti}$ should represent a more accurate measure of IFTA given that it is excluding the glomerular compartment from its calculation. However, as seen in Table 2, the difference in percentages for IFTA versus $IFTA^{ti}$ appear negligible and did not cause any changes to Banff scores for IFTA, which is perhaps not surprising, as glomeruli account for a very small percent of overall cortical volume than tubules.

## 4. DISCUSSION

With progressive technological advances in computing power, speed, and the application of graphic processing units to deep learning systems, it is now feasible to train machine learning algorithms built on CNNs to classify pixel-based image data for various applications with high accuracy. Recently, there has been intense interest in applying this powerful technology to automate image recognition in the medical field, particularly in the diagnostic fields of radiology and pathology. In terms of nephrology, renal pathology seems to be the most logical area of developing and using such image recognition algorithms. Given the possibility of precise quantitation of various morphologic features using deep learning algorithms, research into this field may lead to more effective use of image data derived from biopsies for the assessment of prognosis and prediction of treatment response.

Glomerulosclerosis and IFTA represent chronic changes that are indicative of irrecoverable renal mass. These histopathologic findings, particularly IFTA, are known to be prognostic indicators, independent of disease etiology, and thus, precise quantitation of these features is relevant to essentially all renal biopsies. In this study, we developed

CNN-based algorithms that were trained to detect glomerulosclerosis and IFTA in renal biopsies. Although automated detection of glomerulosclerosis has already been achieved by several groups including our own, to our knowledge this study is the first report of using CNNs for automated detection of IFTA.

Testing various configurations of CNNs revealed that, for IFTA detection, using smaller networks that used low-resolution analysis resulted in superior performance. The smaller networks allowed more training images to be batched at one time, and low-resolution analysis better emulated the pathologist's typical approach to IFTA assessment and provided sufficient contextual information for the CNN in the training set. For detection of non-sclerotic and sclerotic glomeruli, most of the CNNs showed comparable performance, independent of encoder type (i.e. network size) and the resolution of analysis. The overall best performing model developed in this study analyzed the biopsy images using low-resolution with the ResNet-50 encoder. This network demonstrated a balanced accuracy of $0.82/0.94/0.86$ and MCC of $0.6/0.87/0.68$ for localization of IFTA, non-sclerotic glomeruli, and sclerotic glomeruli, respectively, on a mixed intra-institutional holdout set of 14 biopsies consisting of both native kidney biopsies with diabetic nephropathy as well as transplant biopsies that were stained at two separate institutions. Testing the model on an inter-institution holdout set of biopsies (i.e. from an institution that the model was never trained on) showed slightly lower performance with a balanced accuracy of $0.7/0.93/0.84$ and MCC of $0.49/0.79/0.64$ for localization of IFTA, non-sclerotic glomeruli, and sclerotic glomeruli, respectively.

When applying CNNs to biological and medical image data, magnification (i.e. field of view) is a critical parameter that should be selected based on the task at hand and the desired outcome. Since IFTA typically affects regions of the cortex rather than small foci at the individual tubule level, proper IFTA detection by CNNs required a low-resolution view of the data, similar to the pathologist's approach of using lower magnification to estimate the degree of IFTA in a biopsy. Low-resolution analysis provides sufficient contextual information in the training images. To increase CNN performance. In contrast, a low-resolution approach may not be ideal for some tasks that would normal require a pathologist to examine the biopsy at high power. For example, high-resolution analysis would be referred for a task such as nuclei localization, requiring small image sizes and high resolution for optimized CNN performance.

All configurations of CNNs that we tested showed extremely high specificity. Since CNN training is strongly influenced by the number of images that are observed in each class, these high specificities are a direct result of class imbalance in the WSIs. More specifically, there was a much larger number of training images belonging to non-regions of interest (i.e. unscarred renal cortex) than regions of interest (i.e. IFTA, non-sclerotic glomeruli, and sclerotic glomeruli). Therefore, the networks became strongest at identifying the non-regions of interest structures, which leads to high specificity. Similarly, the CNNs showed decreased performance at identifying sclerotic glomeruli compared to non-sclerotic glomeruli, for the same reason; there were less sclerotic glomeruli than non-sclerotic glomeruli in the training set.

There are many studies in the literature which demonstrate techniques which can be used to change the balance of sensitivity and specificity. Most of them depend on balancing the number of training image patches from each class, so that the network does not see one class far more often than others. Another common strategy is to add a bias to the network output probabilities to change the likelihood that the network will segment a specific pixel as a specific class, as we did for nuclear segmentation in our past work[27]. However, in this study, high specificity at the expense of lower sensitivity was preferred since the networks were able to be more precise in its predictions given the imprecise ground truth annotations.

We also compared the pathologist's visual estimations of IFTA on the same cases that he previously annotated, after a washout period, and compared these results to the CNN's predictions. The pathologist's visual estimates were highly correlated to his annotated boundaries (coefficient 0.96 and 95% CI [0.88, 0.98]), and the CNN predictions were moderately to strongly correlated with the annotated boundaries (coefficient 0.73 and 95% CI [0.31, 0.91]). When the percentages of IFTA were converted to Banff scores (ci/ct), there was not a substantial difference in score agreement between the pathologist's visual assessment and his annotations (kappa 0.69, 95% CI [0.39, 0.99]) versus the CNN's predictions and the annotations (0.66, 95% CI [0.37, 0.96] for CNN), indicating that the CNN predictions performed similarly in terms of Banff scoring of IFTA as the pathologist. However, this loss of quantified precision due to categorical scores speaks to the need for the shift of pathology from discrete categorical variables to continuous metrics, which will require computational algorithms to alleviate manual burden and increase consistency.

The application of state-of-the-art convolutional algorithms to medical images still requires careful planning and forethought to achieve ideal performance, despite significant strides being made to increase their accessibility and interpretability. Even in this pilot study, where only cases with clear definition between IFTA and non-IFTA were used, there were considerable engineering decisions that needed to be made to achieve success. However, these approaches provide invaluable access to large scale patterns derived from huge data volumes. Overall, we consider the success of this pilot study an important step towards the next era of precision computational medicine in pathology.

**Author contributions**

BG trained the network models used in this work, performed the quantitative analysis, organized the computational design, and co-wrote the manuscript. KYJ conceived the research scheme, co-wrote the manuscript, performed annotation of ground truth, provided the transplant biopsy data, provided clinical feedback to optimize the design of the computational strategy, and helped conceptualize the work. AR provided ground truth annotation and conceptual feedback. FY selected proper cases for training data from the UC Davis archival records. SJ and ABF provided cases, and gave overall advice and input on the manuscript. PS conceived the research scheme with KYJ and BG, coordinated with the study team, conceptualized the study, critically analyzed the results, and assisted in manuscript preparation.

**Acknowledgments**

The project was supported by the faculty startup funds from the Jacobs School of Medicine and Biomedical Sciences, University at Buffalo, the University at Buffalo IMPACT award, NIDDK Diabetic Complications Consortium grant DK076169, NIDDK grant R01 DK114485, and NIDDK grant U01-DK103225. We thank Ellen Donnert for her assistance in selecting DN biopsies from the VUMC collection. We acknowledge the assistance of the Histology Core Laboratory and Multispectral Imaging Suite in the Department of Pathology & Anatomical Sciences, Jacobs School of Medicine and Biomedical Sciences, University at Buffalo. We thank NVIDIA Corporation for the donation of the Titan X Pascal GPU used for this research (NVIDIA, Santa Clara, CA). We are grateful to Diane Salamon assistance in identifying cases for the study and the support of the Kidney Translational Research Center (KTRC) and the Division of Nephrology at the Washington University School of Medicine in St. Louis, MO.

**Table 1.** Performance of various prediction configurations for localization of IFTA, glomeruli, and sclerotic glomeruli in low, high, and low+high resolution. Intra-institutional holdout cases were produced in the same institutes as the training data, inter-institutional holdout cases were prepared in an institute never before seen by the classifier. Abbreviations: IFTA, interstitial fibrosis and tubular atrophy; NS-G, non-sclerotic glomeruli; GS, glomerulosclerosis; MCC, Matthews correlation coefficient; D, DeepLab; R, ResNet 50; HR, high resolution; LR, low resolution.

| | Sensitivity | | | Specificity | | | Balanced Accuracy | | | MCC | | |
|---|---|---|---|---|---|---|---|---|---|---|---|---|
| | IFTA | NS-G | GS | IFTA | NS-G | GS | IFTA | NS-G | GS | IFTA | NS-G | GS |
| | *Intra-institutional holdout* | | | | | | | | | | | |
| D, LR | 0.6 | 0.89 | 0.73 | 0.99 | 1 | 1 | 0.8 | 0.95 | 0.87 | 0.54 | 0.83 | 0.66 |
| D, HR | 0.34 | 0.92 | 0.55 | 1 | 1 | 1 | 0.67 | 0.96 | 0.77 | 0.46 | 0.86 | 0.53 |
| D, LR+HR | 0.31 | 0.9 | 0.63 | 1 | 1 | 1 | 0.65 | 0.95 | 0.82 | 0.47 | 0.87 | 0.61 |
| R, LR | 0.64 | 0.89 | 0.72 | 0.99 | 1 | 1 | 0.82 | 0.94 | 0.86 | 0.6 | 0.87 | 0.68 |
| R, HR | 0.65 | 0.93 | 0.79 | 0.99 | 1 | 1 | 0.82 | 0.96 | 0.9 | 0.56 | 0.85 | 0.61 |
| R, LR+HR | 0.56 | 0.9 | 0.7 | 1 | 1 | 1 | 0.78 | 0.95 | 0.85 | 0.58 | 0.86 | 0.58 |
| | *Inter-institutional holdout* | | | | | | | | | | | |
| D, LR | 0.32 | 0.89 | 0.67 | 0.99 | 1 | 1 | 0.66 | 0.94 | 0.84 | 0.43 | 0.74 | 0.6 |
| D, HR | 0.22 | 0.92 | 0.61 | 1 | 0.99 | 1 | 0.61 | 0.96 | 0.81 | 0.31 | 0.69 | 0.54 |
| D, LR+HR | 0.1 | 0.89 | 0.54 | 1 | 1 | 1 | 0.55 | 0.94 | 0.77 | 0.21 | 0.76 | 0.6 |
| R, LR | 0.41 | 0.87 | 0.69 | 0.99 | 1 | 1 | 0.7 | 0.93 | 0.84 | 0.49 | 0.79 | 0.64 |
| R, HR | 0.49 | 0.91 | 0.63 | 0.99 | 0.99 | 1 | 0.74 | 0.95 | 0.82 | 0.47 | 0.7 | 0.5 |
| R, LR+HR | 0.28 | 0.86 | 0.56 | 1 | 1 | 1 | 0.64 | 0.93 | 0.78 | 0.41 | 0.75 | 0.55 |

**Table 2. Comparison of human and computer prediction of IFTA. Visual refers to an *in cerebro* accumulation of cortical IFTA percentage by human pathologist. Abbreviations: IFTA, interstitial fibrosis and tubular atrophy; DN, diabetic nephropathy; TX, transplant**

| Case | Visual | | Pathologist annotated boundaries | | | | Computer predicted boundaries | | | |
|---|---|---|---|---|---|---|---|---|---|---|
| | IFTA | Banff IFTA | IFTA | IFTA$^{ti}$ | Banff IFTA | Banff IFTA$^{ti}$ | IFTA | IFTA$^{ti}$ | Banff IFTA | Banff IFTA$^{ti}$ |
| TX 1 | 25.0% | 1 | 23.65% | 24.92% | 1 | 1 | 10.64% | 11.40% | 1 | 1 |
| TX 2 | 35.0% | 2 | 24.10% | 24.44% | 1 | 1 | 15.67% | 16.12% | 1 | 1 |
| TX 3 | 25.0% | 1 | 29.23% | 31.38% | 2 | 2 | 9.78% | 10.66% | 1 | 1 |
| TX 4 | 60.0% | 3 | 53.04% | 51.76% | 3 | 3 | 38.25% | 40.60% | 2 | 2 |
| TX 5 | 30.0% | 2 | 19.83% | 20.06% | 1 | 1 | 17.26% | 18.29% | 1 | 1 |
| TX 6 | 50.0% | 2 | 44.64% | 45.37% | 2 | 2 | 39.18% | 41.19% | 2 | 2 |
| TX 7 | 20.0% | 1 | 20.53% | 19.67% | 1 | 1 | 14.70% | 14.68% | 1 | 1 |
| TX 8 | 40.0% | 2 | 42.00% | 43.76% | 2 | 2 | 29.16% | 31.06% | 2 | 2 |
| TX 9 | 0.0% | 0 | 0.27% | 0.28% | 0 | 0 | 0.68% | 0.71% | 0 | 0 |
| TX 10 | 25.0% | 1 | 16.58% | 17.54% | 1 | 1 | 15.22% | 16.14% | 1 | 1 |
| DN 1 | 15.0% | 1 | 16.87% | 17.53% | 1 | 1 | 36.88% | 38.77% | 2 | 2 |
| DN 2 | 15.0% | 1 | 7.29% | 7.98% | 1 | 1 | 18.03% | 19.25% | 1 | 1 |
| DN 3 | 60.0% | 3 | 56.14% | 56.56% | 3 | 3 | 141.73% | 153.66% | 3 | 3 |
| DN 4 | 5.0% | 1 | 4.19% | 4.31% | 0 | 0 | 5.58% | 5.73% | 1 | 1 |

**Table 3. Comparison of human and computer prediction of glomerulosclerosis. Abbreviations: NS-G, non-sclerotic glomeruli; GS, glomerulosclerosis; DN, diabetic nephropathy; TX, transplant.**

|        | Pathologist annotated glomeruli | | | Computer predicted glomeruli | | |
|--------|------|-----|--------|------|-----|--------|
| Case   | NS-G | GS  | % GS   | NS-G | GS  | % GS   |
| TX 1   | 44   | 0   | 0.00%  | 44   | 0   | 0.00%  |
| TX 2   | 8    | 1   | 11.11% | 8    | 1   | 11.11% |
| TX 3   | 33   | 5   | 13.16% | 33   | 5   | 13.16% |
| TX 4   | 12   | 19  | 61.29% | 15   | 18  | 54.55% |
| TX 5   | 46   | 0   | 0.00%  | 45   | 4   | 8.16%  |
| TX 6   | 35   | 17  | 32.69% | 28   | 19  | 40.43% |
| TX 7   | 55   | 12  | 17.91% | 44   | 13  | 22.81% |
| TX 8   | 19   | 2   | 9.52%  | 19   | 4   | 17.39% |
| TX 9   | 23   | 0   | 0.00%  | 21   | 0   | 0.00%  |
| TX 10  | 10   | 1   | 9.09%  | 9    | 0   | 0.00%  |
| DN 1   | 11   | 0   | 0.00%  | 9    | 0   | 0.00%  |
| DN 2   | 14   | 12  | 46.15% | 15   | 14  | 48.28% |
| DN 3   | 5    | 4   | 44.44% | 5    | 5   | 50.00% |
| DN 4   | 4    | 1   | 20.00% | 4    | 1   | 20.00% |

**Figure 1**

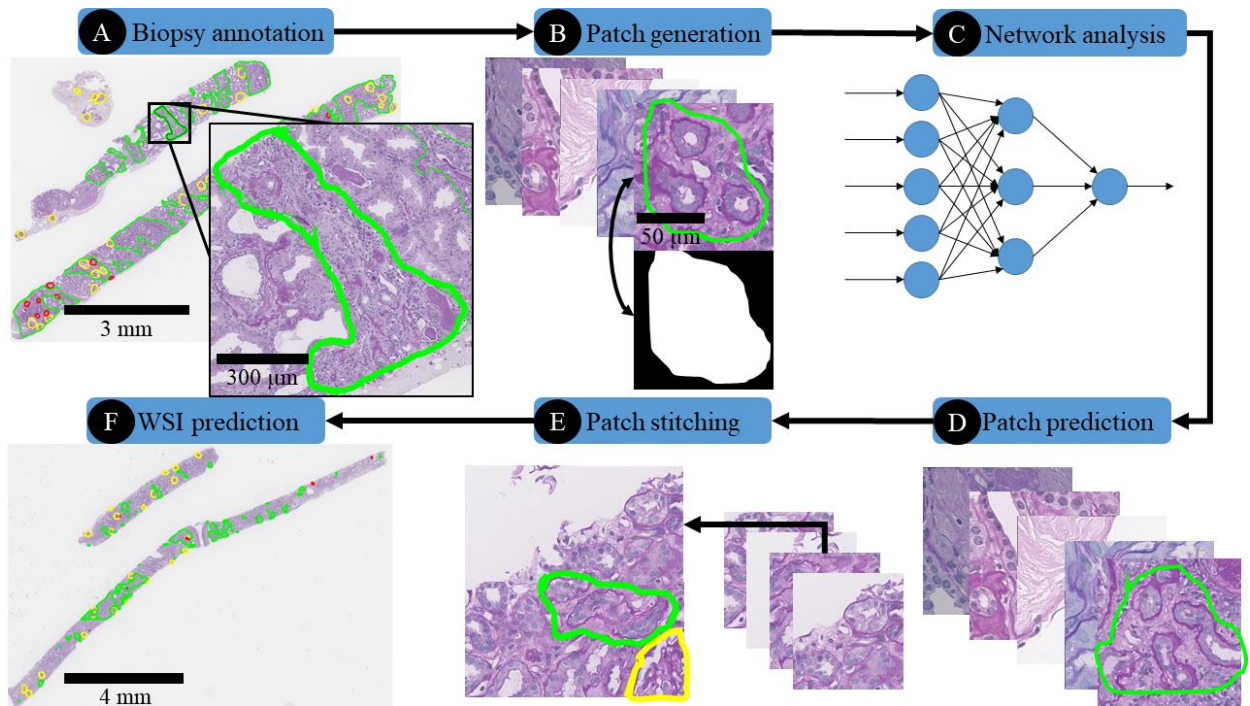

**Figure 1. Schematic overview of convolutional neural network (CNN) analysis using the Human Artificial Intelligence Loop (H-AI-L) pipeline.** (**A**) Biopsies were annotated by a pathologist who circled regions of interest; as an example, IFTA is shown circled in green. (**B**) Whole slide images were chopped into image patches and masks were created for each image patch based on the annotations. (**C**) Image patches and their corresponding masks were used to train a CNN model. (**D**) The CNN model was used to predict regions of interest on test image patches, which were represented by prediction masks. (**E**) Test images and their corresponding prediction masks were stitched back together. (**F**) Prediction masks were converted to XML format to display boundaries of CNN-predicted regions of interest on whole slide images.

**Figure 2**

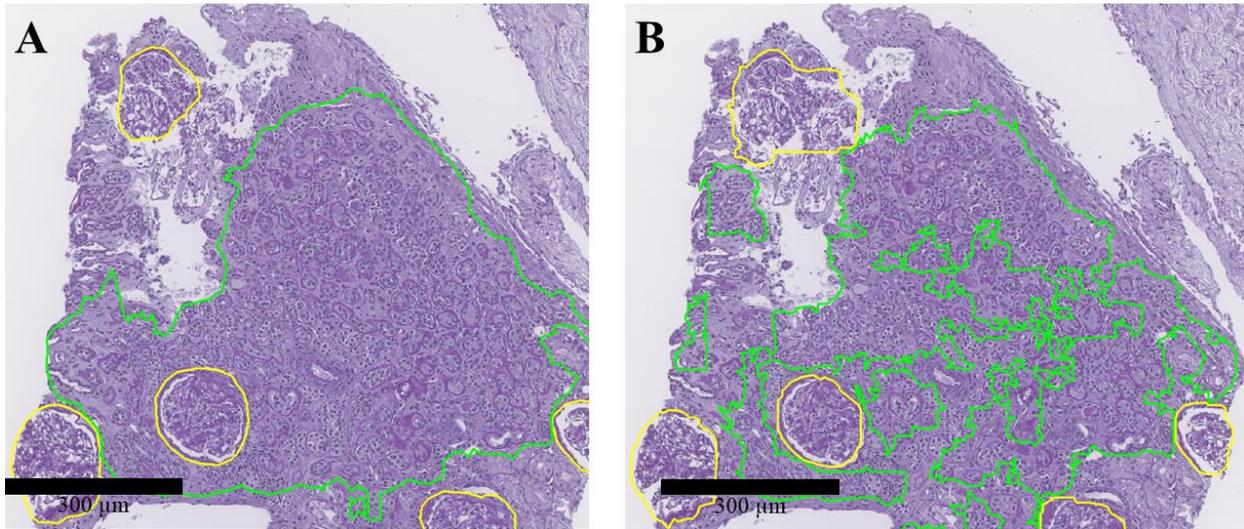

**Figure 2. Comparison of aggregated interstitial fibrosis and tubular atrophy (IFTA) predictions in low and high resolution approaches.** Low resolution network approach (**A**) aggregates large regions of IFTA (green boundaries) more effectively than a high resolution approach (**B**). The output of the low resolution approach is more similar to the typical renal pathologist's approach to IFTA estimation performed at low magnification.

Figure 3

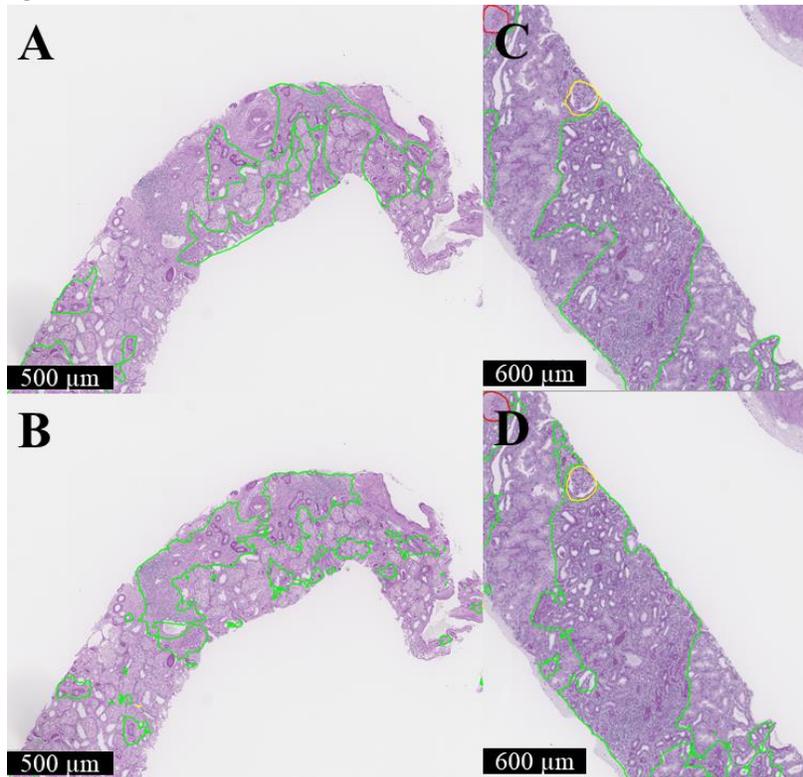

**Figure 3. Comparison of patch-level overlap with ground truth for high and low Matthew's correlation coefficient (MCC).** **(A)** Pathologist annotations for a transplant holdout case that scored lower than average MCC value. **(B)** Prediction image for (A) which scores MCC of 0.51. **(C)** Pathologist annotations for a transplant holdout case which scored higher than average MCC value. **(D)** Prediction image for (C) which scores an MCC of 0.84.

**Figure 4**

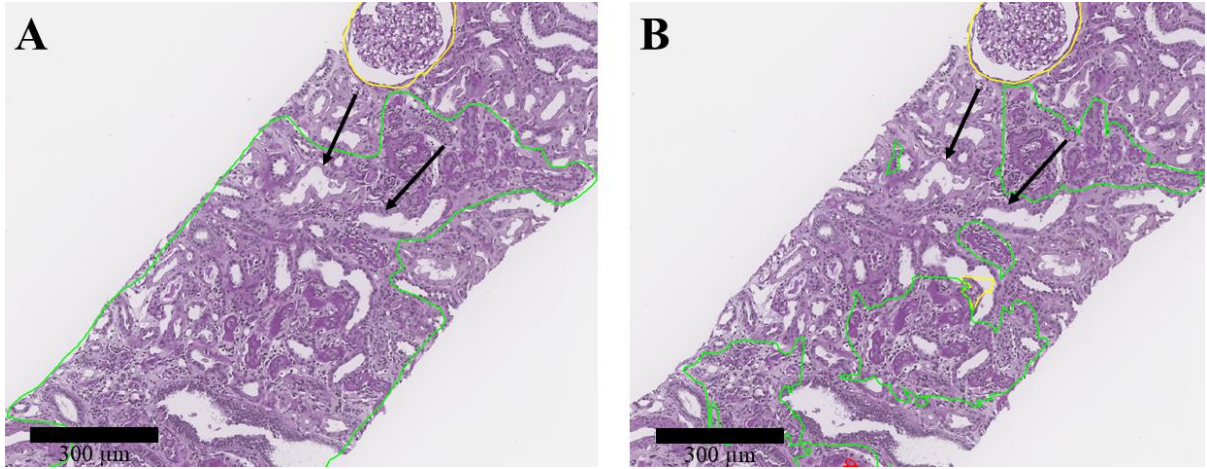

**Figure 4. Example of network generalization beyond annotated ground truth.** In this holdout biopsy **(A)** annotated ground truth was generated by the pathologist by broadly circling regions of interstitial fibrosis and tubular atrophy (IFTA) at low magnification. **(B)** Network prediction of IFTA regions which demonstrate the avoidance of non-atrophic tubules, obfuscating performance analysis.

**Figure 5**

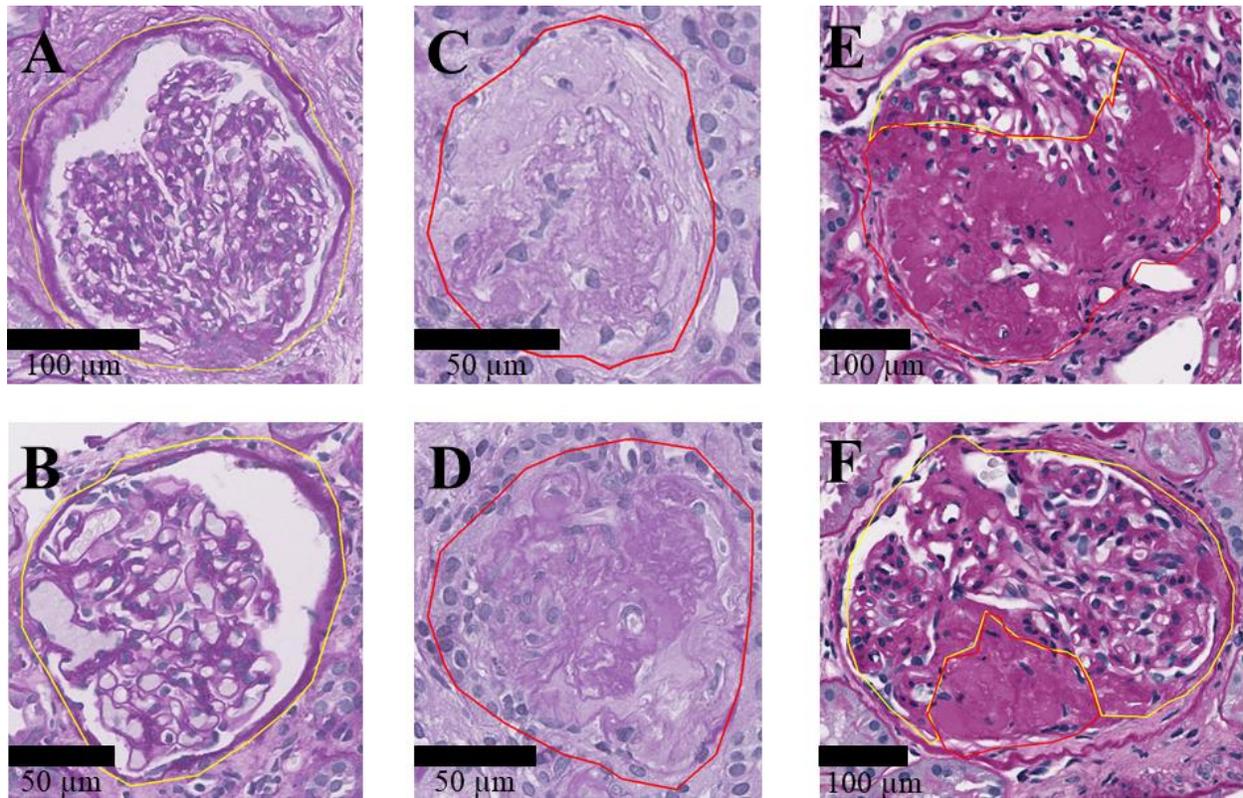

**Figure 5. Network predictions for classes of glomeruli.** Yellow boundaries indicate non-sclerotic glomerulus class and red boundaries indicate sclerotic glomerulus class. **(A, B)** Network boundary prediction for non-sclerotic glomeruli. **(C, D)** Network boundary prediction for sclerotic glomeruli. **(E, F)** Network self-learned segmental glomerulosclerosis identification.

**Figure 6**

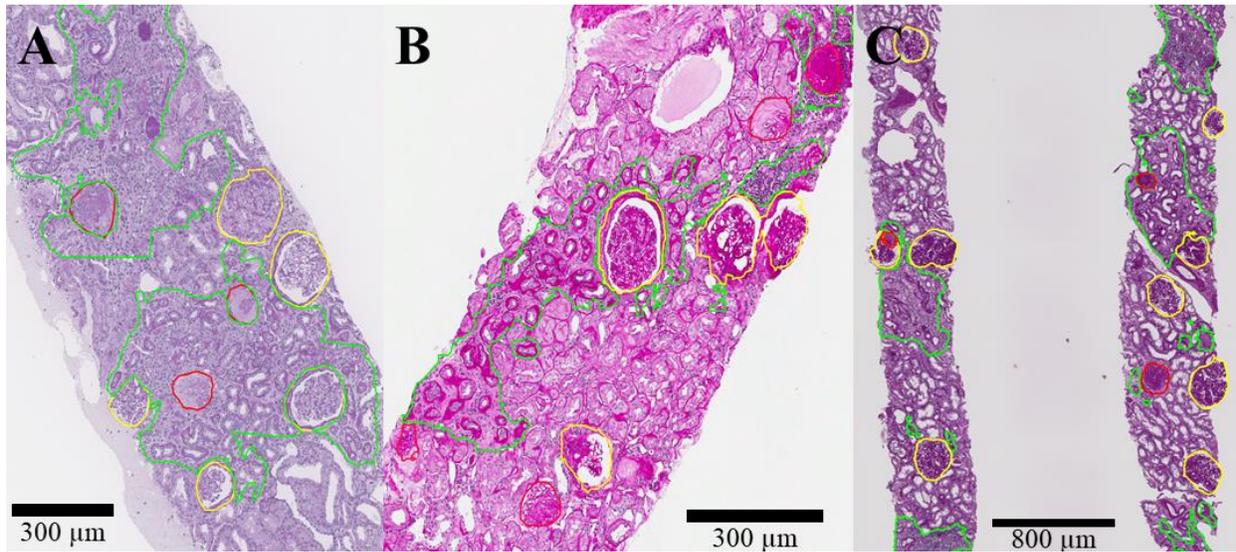

**Figure 6. Examples of convolutional neural network (CNN) segmentation performance in three different PAS stain variations produced by three different institutions.** Examples of interstitial fibrosis and tubular atrophy (green boundaries), non-sclerotic glomerulus (yellow boundaries), and sclerotic glomerulus (red boundaries) predictions by the CNN in: **(A, B)** intra-institutional holdout whole slide images (WSIs) and **(C)** an inter-institutional holdout WSI. Intra-institutional holdout WSIs consisted of cases that were produced at the same institution as the data that the CNN was trained on, while inter-institutional holdout WSIs consisted of cases that were produced in a different institution than the data on which the CNN was trained.

**Supplementary Figure 1. Erroneous prediction of medulla as renal cortical interstitial fibrosis and tubular atrophy.** The convolutional neural network mistakenly classified medulla consisting of small caliber tubules with thickened, reduplicated basement membranes and hyaline casts as interstitial fibrosis and tubular atrophy within the cortex (green boundaries).

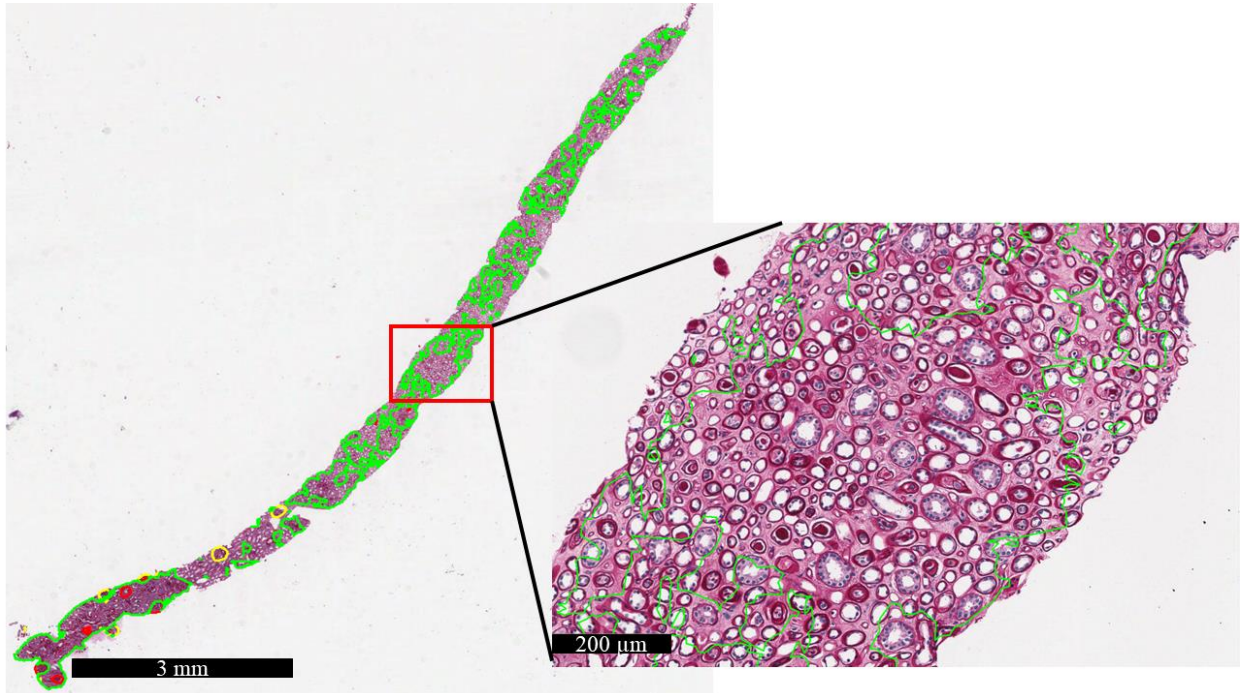